\documentclass[twocolumn]{aastex7}
\usepackage{amsmath}
\usepackage{float}
\usepackage{tikz}
\usepackage{natbib}

\newcommand{\parfrac}[2]{\left(\frac{#1}{#2}\right)}

\newcommand{\Ns}{N_\star}
\newcommand{\trlx}{\tau_{\rm rlx}}

\begin{document}

\title{The $M$-$\sigma$ Relation Has To Break}

\author[orcid=0009-0001-0346-6434]{Omri Nitzan}
\affiliation{Racah Institute of Physics, The Hebrew University of Jerusalem,  9190401, Israel}
\email[show]{omri.nitzan@mail.huji.ac.il}

\author[orcid=0000-0002-1084-3656]{Reem Sari}
\affiliation{Racah Institute of Physics, The Hebrew University of Jerusalem,  9190401, Israel}
\email{sari@phys.huji.ac.il}

\correspondingauthor{Omri Nitzan}
\begin{abstract}

We revisit the growth of central black holes via tidal disruption events (TDEs) and plunges of stellar-mass black holes (sBHs). Our model incorporates the current understanding of mass segregation, where sBHs sink to the center, enhancing the rates of both TDEs and plunges.  

We demonstrate that in dense cluster cores, with densities exceeding $10^6\,\mathrm{M_\odot\,{pc}^{-3}}$,  seeds of initial mass \(M_0\gtrsim100\,M_\odot\) undergo runaway growth. This runaway terminates once the black hole radius of influence surpasses the core radius, or equivalently, most of the core mass has been consumed. This typically results in an intermediate-mass black hole (IMBH), within a few \(\mathrm{Gyr}\). Subsequent growth proceeds as a power law. 

In contrast to observed supermassive black holes (SMBHs), which tightly follow the famous $M \propto \sigma^\beta$ with $\beta \cong5$, our derived sBH accretion rates, integrated over a galactic lifetime, predict final masses of  \(M\approx10^5\,M_\odot \times(\sigma/50\mathrm{km\;s^{-1}})^{2.5}\). 
While our prediction for the contribution of plunges to the growth of the IMBH is robust, the TDE contribution can be negligible if only a small fraction of their mass is actually accreted or up to 3 times higher than the plunges contribution if half a stellar mass gets accreted in each event. Below \(M\sim10^5\,M_{\odot}\) this accreted star and sBH mass exceeds the extrapolation of the observed $M$-$\sigma$ relation. This predicts that the $M$-$\sigma$ scaling must flatten below \(M \sim 10^5\,M_{\odot}\) to a shallower, $2.26<\beta<2.5$ profile.
If confirmed by observations, this would indicate capture-driven growth. 
\end{abstract}
\keywords{
  \uat{Tidal disruption}{1696} ---
  \uat{Intermediate-mass black holes}{816} ---
  \uat{Supermassive black holes}{1663} ---
  \uat{Globular star clusters}{656} ---
  \uat{Stellar dynamics}{1596} ---
  \uat{Galaxy nuclei}{609} 
}

\section{Introduction}\label{sec:intro}
Supermassive black holes (SMBHs) with masses \(M\sim10^{6}M_\odot-10^{10}\,M_\odot\) reside at the centers of most galaxies and power the active galactic nuclei (AGN) episodes that dominate cosmic accretion history \citep{1998AJ....115.2285M,2004MNRAS.351..169M}. The prevailing SMBH growth paradigm combines gas accretion with hierarchical black hole (BH) mergers \citep{2003ApJ...596L..27K}. An increasing body of evidence points to a population of IMBHs (\(10^{3}-10^{6}\,M_\odot\)) in dwarf galaxies, nuclear star clusters and globular clusters \citep{L_tzgendorf_2013,2006ApJ...641L..21G}. How such seeds form and evolve remains an open question. 

A cornerstone observational result is the tight $M$-$\sigma$ relation linking SMBH mass to host bulge stellar velocity dispersion \citep{2000ApJ...539L...9F,Kormendy_2013}. We use the fit by \cite{McConnell_2013}
\begin{equation}
    M =  1.9\times10^8M_{\odot}\cdot\parfrac{\sigma}{200 \mathrm{{km\,s^{-1}}}}^{5.1}.
    \label{eq:M_sigma}
\end{equation}
While nearly linear in \(\log M-\log\sigma\) for SMBH (above $10^6M_{\odot}$), recent work finds a flattening or increased scatter below \(M\sim10^6M_{\odot}\) \citep{2006ApJ...641L..21G,dwarf_galaxies_M_sigma,Xiao_2011}.
The behavior of this scaling at low velocity dispersion is crucial for distinguishing growth models and clarifying how IMBHs connect to the SMBH population.

We examine the capture of stars and stellar mass BHs (sBHs) as a growth mechanism of massive BH (MBH). Mutual interactions cause the orbits of stars and sBHs to diffuse into the black hole’s loss cone \citep{Shapiro1977ApJ,1976MNRAS.176..633F,Magorrian_1999}. sBHs are consumed whole, while stars are tidally disrupted (TDEs), depositing some fraction of their mass. Observed TDE rates of \(\sim10^{-4}-10^{-5}\,\mathrm{yr^{-1}}\) per galaxy \citep{2018ApJ...852...72V}, together with the weak dependence on mass \citep{Magorrian_Tremaine_Scott_1999} 
suggest that capture-driven feeding over a period of $10\, \mathrm{Gyr}$ may lead to accretion of order $10^5M_\odot$. Therefore, as already suggested by \citet{Nicholas_runawayBH_2017}, this could be the dominant growth mechanism for low MBH masses, or low \(\sigma\).

In this work we derive the rate of black hole growth powered solely by stellar and sBH capture using loss cone theory. Our model assumes the MBH is stationary and surrounded by a nuclear stellar cluster. 
The cluster is modeled as a population of $1\,M_\odot$ stars and $10\,M_\odot$ sBHs. We derive the mass segregation between these two populations using the constant energy flux, as in \cite{rom2025segregationnuclearstellarclusters,Linial_2022}. In these models, most stars are at the outskirts of the sphere of influence, while most of the BHs are at distances shorter by an order of magnitude.

We derive closed‐form expressions for the rates at which both populations fall into the loss cone taking into account empty and full loss cone regimes
\citep{Shapiro1977ApJ,1976MNRAS.176..633F,Magorrian_1999}.

We assume the surrounding cluster maintains a constant \(\sigma\) outside the sphere of influence of the MBH, but allow the MBH to influence the density inside it. We integrate the mass growth over time. We demonstrate that stellar and sBH feeding drive runaway growth for seeds exceeding some threshold into an IMBH consisting of most of the initial cluster core mass.

This runaway terminates once the cluster core is consumed. Growth of the MBH then continues more slowly, roughly as square root of time. We then predict that the present day MBH masses follow \(M\propto\sigma^{\beta}\), with $2.26<\beta<2.5$. This curve intersects the extrapolation of the observed $M$-$\sigma$ relation \eqref{eq:M_sigma} at \(\sigma\sim50\,\mathrm{km\,s^{-1}}\), equivalent to $M\sim10^5M_\odot$. We therefore predict a break in the known $M$-$\sigma$ relation below
\(\sigma\sim50\,\mathrm{km\,s^{-1}}\).

The paper is organized as follows: Section \ref{sec:setup} describes our setup; Section \ref{sec:twoMass} calculates capture rates in the segregated two-mass model, and then integrates the derived rates to find expected present day masses;  Section \ref{sec:GC Core} applies the theory to cluster cores and quantifies the seed masses required for runaway growth; Finally, we summarize our findings in Section \ref{sec:conclusions}.
\section{Setup}
\label{sec:setup}
\begin{figure}
    \centering
    \includegraphics[width=1\linewidth]{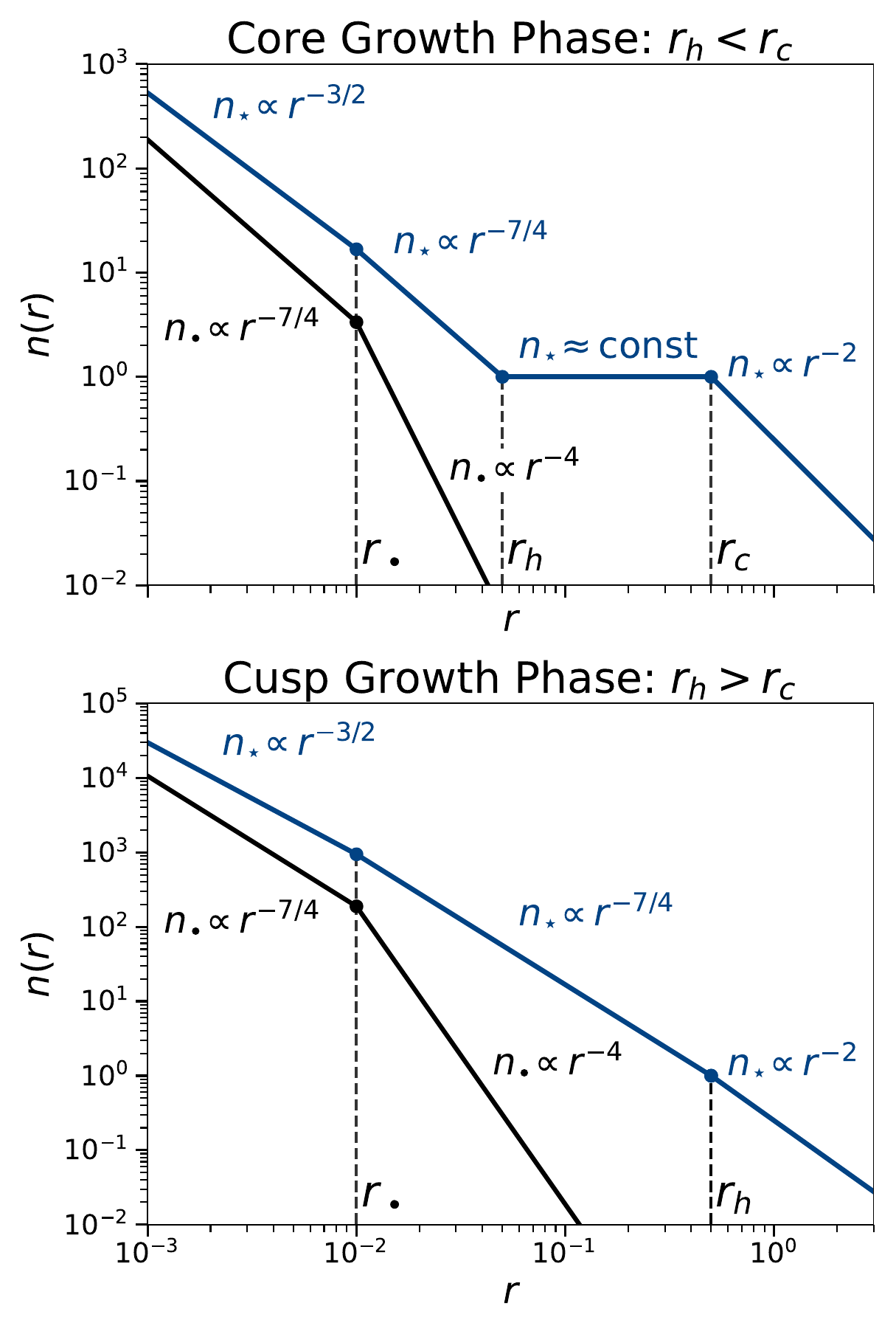}
    \caption{Predicted density of sBH (black) and stars (blue) in the cluster in both growth phases}
    \label{fig:density}
\end{figure}

Observed clusters consist of a uniform density core up to some radius \(r_c\), beyond which the density follows a cusp profile \citep{1966AJ.....71...64K,binney2008galactic}. We model this as an isothermal sphere with a core of constant density and velocity dispersion (\(\rho_c, \sigma\))
\begin{equation}
\rho=
\begin{cases}
\rho_c \;\;\;\;\;\;\;\;  r<r_c \\ \rho_{\mathrm{ IS}} =\frac{\sigma^2}{2\pi G r^2}  \;\;\;r>r_c  \end{cases}.
\end{equation}
Using the virial theorem, we estimate \(r_c\)
\begin{equation}
    r_c = \frac{\sigma}{\sqrt{\frac{4}{3}\pi G \rho_c}}.
\end{equation}
 The presence of a massive black hole modifies the density profiles within its radius of influence, defined as  \(r_h ={GM}/{\sigma^{\,2}}\), where the MBH dominates the gravitational potential.
 Within the radius of influence, the density is described by the segregated two-mass Bahcall-Wolf (BW) cusp \citep{1977BW,Alexander_2009_Storng_SEGREGATION, Linial_2022}. We distinguish two cases: the radius of influence is larger than the core radius, which we call a cusp profile, or smaller than the core radius, which we call core profile (see Figure \ref{fig:density}). 
 
\subsection{Mass Segregation}
\label{subsec:Massseg}
In the following derivation, we calculate the relaxed density profile for stars and sBHs in the two-mass BW cusp.
We use a two-mass model with \(m_{\star} = 1M_{\odot}\) stars and \(m_{\bullet} = 10M_{\odot}\) sBH. Dynamical mass segregation sinks sBHs down to a characteristic radius \(r_{\bullet}\). In a dynamically relaxed environment, the steady-state distribution satisfies a constant energy flux \citep{1977BW,Fragione_2018,rom2025segregationnuclearstellarclusters}
\begin{equation}\label{eq:Eflux}
    \mathcal{F}_E\approx \frac{E(r,m)\Ns(m)}{\trlx(r,m)}.
\end{equation}
Using the relaxation time \eqref{eq:Trel} we find the energy flux scales as $\mathcal{F}_E\propto m^3n^2r^{-5/2}$ (\(n\) is the number density of objects of mass \(m\)). At \(r_\bullet\) the dominant scatterers switch from sBH to stars, we can therefore equate the energy flux from star and sBH scatterings and get
\begin{equation}
\label{eq:numdens}
    \left. \frac{n_{\star}}{n_\bullet}\right|_{r=r_\bullet} = \parfrac{m_{\star}}{m_{\bullet}}^{3/2}.
\end{equation}
In their work \cite{1977BW} have found that the density profile of objects of mass \(m\) scattered predominantly by objects of mass \(m_{dom}\) follows \(n(r)\propto r^{-(3/2+m/4m_{dom})}\). 
Therefore, for \(r<r_\bullet\) sBHs maintain a Bahcall–Wolf profile with \(n_\bullet\propto r^{-7/4}\) while stars follow a shallower \( n_\star\propto r^{-3/2}\) profile . Beyond  \(r_\bullet\) stars follow \(n_\star\propto r^{-7/4}\) and the sBH population steepens rapidly and becomes dynamically negligible.\\
We integrate these densities to find the total number of sBHs and stars in the sphere of influence. Setting \(f_\bullet = N_{\bullet}/N_{\star}\) as the sBH fraction in the sphere of influence we get
\begin{equation}
    r_{\bullet} \approx \left(f_\bullet^2\parfrac{m_{\bullet}}{m_{\star}}^3\right) ^{2/5}r_h.
\end{equation}
Notice that $f_\bullet$ represents the fraction of sBHs specifically within the sphere of influence. This value is enhanced compared to the global galactic sBH fraction as massive compact remnants segregate toward the center. \cite{rom2025segregationnuclearstellarclusters} estimate that $f_{\bullet} \approx 3f_{\text{global}}$, this enhancement factor may be even larger for the smaller galaxies of interest here. 

The normalization of the densities is chosen such that the total mass in stars equals the mass of the MBH in the cusp case, while in the core case the density at $r_h$ is given by the core density. 
Consequently, the sBH density is found using \eqref{eq:numdens}, i.e. their number density at $r_\bullet$ is smaller than the stellar density by a factor of $(m_\bullet/m_\star)^{3/2}\cong 30$.

\subsection{Capture Rates}
Using loss cone theory we can find the capture rate under the empty loss cone and full loss cone approximations \citep{1976MNRAS.176..633F,Shapiro1977ApJ}. 

The empty loss cone approximation holds if \(\Delta J\), the change in angular momentum over one orbit, is small compared to the loss cone angular momentum \(J_{LC}\) . Unlike the familiar case around SMBHs where the loss cone for TDEs is empty throughout the radius of influence, for IMBHs, empty loss cone occurs only for objects at small radii.

If \(\Delta J\) is large compared to \(J_{LC}\) the full loss cone approximation holds. This is typical of objects at large radii. At some radius \(\Delta J\) equals \(J_{LC}\) we mark it as \(r_d^\star\) for stars and \(r_d^\bullet\) for sBHs. \(r_d\) is the dominant radius from which most infalling objects originate through diffusion in angular momentum.

We mark the capture rate of objects at radius \(r\) per unit \(\log(r)\) as \(\Gamma(r)\) and the rates under the approximations as \(\Gamma_{emp}(r) ,\;\Gamma_{full}(r)\) . The overall rate for a MBH of mass \(M\) is
\begin{equation}
   {\cal R} = \int\Gamma(r)d[\log(r)].
\end{equation}
At the transition radius \(r_d\), rates under both approximations are equal: \(\Gamma_{emp}(r_d) = \Gamma_{full}(r_d)\approx\Gamma(r)\). Since \(r_d\) is the dominant radius we can use  \({\cal R} = \Gamma_{emp}(r_d) \) (given that \(r_d<r_h\) which is the case for the IMBH discussed in this paper).
We estimate the MBH mass accretion rate as
\begin{equation}
    \frac{dM}{dt}=\alpha \, m_\star\,{\cal R_\star} + m_\bullet\,{\cal R_\bullet},
    \label{eq:growth}
\end{equation}
With \(\alpha\) being the fraction of the stellar mass that is accreted in each TDE.

We repeat the standard derivation of the capture rates in a one-mass toy model for our cusp and core models in Appendix \ref{sec:oneMass}. 
To obtain an estimate of the mass capture rate in the two-mass model, we correct the one-mass model rates for the distinct capture efficiencies and loss cone radii of each population.
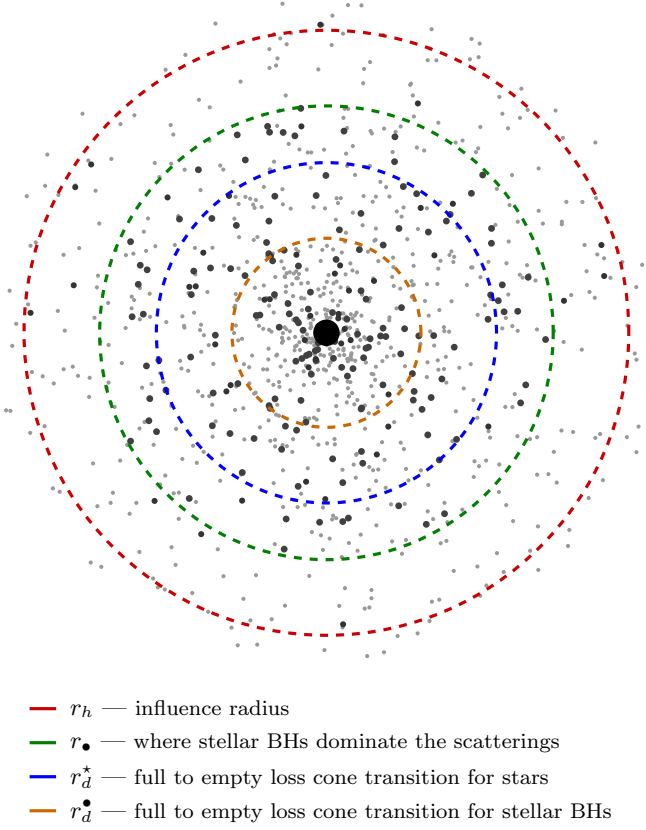
\begin{figure}[t]
    \centering
    \begin{tikzpicture} [scale=5]
  \def\rh{0.8}          
  \def\rbullet{0.6}     
  \def\rd{0.45}         
  \def\rdbullet{0.25}   

  \pgfmathsetseed{134}

        
\foreach \i in {1,...,800}{
    \pgfmathparse{rnd}             \let\U\pgfmathresult
    \pgfmathparse{1.1*\rh*pow(\U,1.2)} \let\r\pgfmathresult
    \pgfmathparse{rnd*360}         \let\a\pgfmathresult
    \pgfmathparse{\r*cos(\a)}      \let\x\pgfmathresult
    \pgfmathparse{\r*sin(\a)}      \let\y\pgfmathresult
    \fill[black!40] (\x,\y) circle (0.0055);
  }
  \pgfmathsetseed{135}
  \foreach \i in {1,...,220}{
    \pgfmathparse{rnd}             \let\U\pgfmathresult
    \pgfmathparse{1.1*\rh*pow(\U,1.5)} \let\r\pgfmathresult
    \pgfmathparse{rnd*360}         \let\a\pgfmathresult
    \pgfmathparse{\r*cos(\a)}      \let\x\pgfmathresult
    \pgfmathparse{\r*sin(\a)}      \let\y\pgfmathresult
    \ifdim \r pt < \rbullet pt
      \fill[black!75] (\x,\y) circle (0.009);      
    \fi
    }

\foreach \i in {1,...,50}{
    \pgfmathparse{rnd}             \let\U\pgfmathresult
    \pgfmathparse{1.1*\rh*pow(\U,1.2)} \let\r\pgfmathresult
    \pgfmathparse{rnd*360}         \let\a\pgfmathresult
    \pgfmathparse{\r*cos(\a)}      \let\x\pgfmathresult
    \pgfmathparse{\r*sin(\a)}      \let\y\pgfmathresult
   \fill[black!80] (\x,\y) circle (0.008);      
  }
  \fill[black] (0,0) circle(0.035);

  \draw[red!80!black , very thick , dashed] (0,0) circle (\rh);        
  \draw[green!50!black, very thick , dashed] (0,0) circle (\rbullet);  
  \draw[blue , very thick , dashed] (0,0) circle (\rd);               
  \draw[orange!80!black , very thick , dashed] (0,0) circle (\rdbullet); 




\end{tikzpicture}
\begin{center}
\begin{tikzpicture}[baseline]

  \draw[red!80!black, very thick] (0,0) -- (0.35,0);
  \node[anchor=west] at (0.42,0)
      {\(\displaystyle r_h\) — \footnotesize\text{influence radius}};

  \draw[green!50!black, very thick] (0,-0.45) -- (0.35,-0.45);
  \node[anchor=west] at (0.42,-0.45)
      {\(\displaystyle r_{\bullet}\) — \footnotesize\text{where stellar BHs dominate the scatterings}};

  \draw[blue, very thick] (0,-0.90) -- (0.35,-0.90);
  \node[anchor=west] at (0.42,-0.90)
      {\(\displaystyle r_d^{\star}\) — \footnotesize\text{full to empty loss cone transition for stars}};

  \draw[orange!80!black, very thick] (0,-1.35) -- (0.35,-1.35);
  \node[anchor=west] at (0.42,-1.35)
      {\(\displaystyle r_d^{\bullet}\) — \footnotesize\text{full to empty loss cone transition for stellar BHs}};

\end{tikzpicture}
\end{center}
  \caption{Schematic of a two-mass Bahcall-Wolf cusp around a central MBH. Mass segregation drives stellar mass BHs (black dots) to the center and stars (gray dots) out of the center. Given the range of parameters of interest, namely $M<10^6M_\odot$, $\sigma<100km/s$, $f_\bullet \sim 10^{-3}$, we obtain $r_h>r_\bullet \gtrsim r^*_d > r^\bullet_d$ as shown in the figure.}
    \label{fig:enter-label}
\end{figure}

\section{Growth in a Cusp Profile}
\label{sec:twoMass}

We derive the mass capture rate for sBH and star capture, then integrate both rates to establish the $M$-$\sigma$ scalings. The growth from sBHs provides a lower limit, while the growth by stars, if we assume $\alpha=1/2$ represent an upper bound on the accreted mass.

\subsection{sBH Mass Capture}
sBH capture occurs if the specific angular momentum falls below the marginally bound (MB) orbit angular momentum \citep{Shapiro1977ApJ,1983Shapiro.book}:
 \begin{equation}
 \label{eq:sBH_par}
    J_{MB} =\frac{4GM}{c}.
\end{equation}

For relevant parameters \(r_d^{\bullet}<r_\bullet\) , therefore sBHs are mostly scattered by other sBHs into the loss cone.
At $r_d^\bullet$, the capture is by a plunge onto the MBH (equation \ref{eq:sBH_par}). Capture following an inspiral by the emission of gravitational waves, known as EMRIs (e.g. \cite{Amaro_Seoane_2018}), occurs at slightly lower radii. While for SMBH EMRIs are hundreds of times less frequent than plunges \citep{rom2025segregationnuclearstellarclusters} for IMBH the ratio is less extreme, yet we still neglect their subdominant contribution.
From \eqref{eq:numdens}, sBH mass density differs from the single mass case by a factor of 
\((m_{\star}/{m_{\bullet}})^{1/2}\). We substitute \(\rho_h = (m_{\star}/{m_{\bullet}})^{1/2}\rho_{IS}(r_h)\) and \eqref{eq:sBH_par} into \eqref{eq:transition_rate} to get the sBH capture rate
\begin{align}
 \label{eq:sBH_rate}
   &{\cal R}_{\bullet}= K_{\bullet}\,M^{-5/9}\\&=7.75\times10^{-7} \mathrm{\mathrm{yr^{-1}}} \parfrac{M}{10^5M_{\odot}}^{-5/9}\parfrac{\sigma}{50\mathrm{km\,s^{-1}}}^{35/9} \nonumber,
\end{align}
where
\begin{equation}
K_{\bullet} =\frac{64}{5^{14/9}\,\pi }\frac{ \sigma^{35/9} }{\ln(\Lambda)^{4/9}m_{\bullet}^{4/9} G c^{8/9}}\parfrac{m_{\star}}{m_{\bullet}}^{7/9} \nonumber.
\end{equation}

We integrate this rate over time starting with \(M(t=0) = 0\) and get the mass as a function of time for a MBH that grows via sBH capture only
\begin{align}
  \label{eq:sBH_mass}
    M_{\bullet}(t) &= \Bigl(\tfrac{14}{9}m_{\bullet}K_{\bullet}\,t\Bigr)^{9/14}\\&\approx  1.13\times 10^5\,M_\odot \left(\frac{t}{10 \mathrm{Gyr}}\right)^{9/14} \left(\frac{\sigma}{50\mathrm{km\,s^{-1}}}\right)^{5/2} \nonumber.
\end{align}
\subsection{TDE Mass Capture}
For TDEs we have
\begin{equation}
\label{eq:TDE_par}
    R_{LC} = \Bigl(\frac{M}{m_{\star}}\Bigr)^{1/3}r_{\star} .
\end{equation}
For relevant parameters \(r_d^{\star}<r_\bullet\) , therefore stars are mostly scattered into the loss cone by sBHs. 

To estimate the TDE rate originating from \(r_d^{\star}\) we start with the sBH capture rate and add factors that correct for difference in parameters. 
The two changes from the sBH rate are a different \(r_d\) and a different density. The stellar density follows
\begin{equation}
      \left. \frac{n_\star}{n_\bullet} \right|_{r=r_d^\star}= \parfrac{m_{\bullet}}{m_{\star}}^{3/2}\parfrac{r_d^{\star}}{r_{\bullet}}^{1/4}.
\end{equation}
The capture rate increases linearly with density since the density of scatterers (sBHs) is left unchanged. \(r_d^{\star}\) is bigger than \(r_d^\bullet\) by a factor of \(\left(R_{LC}^{\star}/R_{LC}^{\bullet}\right)^{4/9}\) . The capture rate increases by the same factor since the empty loss cone rate scales linearly with radius. We get
\begin{align}
 \label{eq:TDE_rate_rd}
      &{\cal R}_{\star}=K_{\star}M^{-22/27}\\ &= 1.13\times10^{-4} \mathrm{yr^{-1}} \parfrac{M}{10^5M_{\odot}}^{-22/27}\parfrac{\sigma}{50\mathrm{km\,s^{-1}}}^{37/9}\nonumber,
\end{align}
where
\begin{equation}
   K_{\star} =\frac{2^{13/3}}{5^{13/9}\pi\ln(\Lambda)^{5/9}}\frac{\sigma^{37/9}r_\star^{5/9}}{m_\bullet^{7/90}m_\star^{179/270}G^{14/9}f_\bullet^{1/5}}.\nonumber
\end{equation}
In \eqref{eq:TDE_rate_rd} we set $f_\bullet=0.002$, but the results weakly depends on this choice. 

By coincidence, the TDE rate for MBHs obeying the observed $M$-$\sigma$  relation (eq. \eqref{eq:M_sigma}) is $\propto M^{37/45-22/27}$, i.e. nearly independent of mass (see also \cite{Chang_2025_rates_of_TDE}).
For masses $M \sim 10^5-10^6\,M_\odot$, where the loss cone is full near the radius of influence ($r_h$), the rate plateaus at approximately $10^{-4}\,{\rm yr}^{-1}$. At lower masses, the rate grows with $M$ as the critical radius $r_d$ moves outward. Conversely, for $M \gtrsim 10^6\,M_\odot$, the loss cone becomes empty even at $r_h$ and the TDE rate decreases slowly. Across all regimes, the segregated sBHs enhance the TDE rate: although they slightly reduce the stellar density, they shorten the relaxation timescale, resulting in a net enhancement of the TDE rate compared to a purely stellar system.

\subsubsection{TDE efficiency }
To estimate the TDE mass accretion rate we need the TDE mass accretion efficiency \(\alpha\),  which is bound between \(0.5\) (all the bound mass is accreted) and the Eddington-limited efficiency (see Appendix \ref{app:EddEffderivation} for the derivation)
\begin{equation}
\alpha_{Edd} = 0.021\,
\parfrac{M}{10^5M_\odot}^{3/5}.
\end{equation}
As we show below, if TDE accretion is Eddington limited it is insignificant compared to the sBH capture. Even though there are significant indications that super Eddington accretion occurs at early stages of the TDE \citep{Krolik_2016,2018ApJ_IMBH_TDE}, the fraction \(\alpha\) of accreted mass for TDEs around smaller mass IMBH is less certain. 

We adopt an insignificant seed mass,\(M(t=0) = 0\), and solve \eqref{eq:growth} with \(\alpha = 0.5\) for growth driven only by TDEs at rate \eqref{eq:TDE_rate_rd}. The MBH mass that grows via TDEs only is
\begin{align}
\label{eq:TDE_mass}
    M_{\star}(t) &= \Bigl(\tfrac{49}{27}\alpha\, m_{\star}\,K_{\star}\,t\Bigr)^{27/49}\\&\approx 3.61 \times 10^5\,M_\odot \left(\frac{t}{10 \mathrm{Gyr}}\right)^{27/49} \left(\frac{\sigma}{50\mathrm{km\,s^{-1}}}\right)^{111/49} \nonumber.
\end{align}
We therefore argue that the total mass accretion rate on MBHs, as given by equation \eqref{eq:growth}, is bound from above by the TDE mass accretion rate for \(\alpha =0.5\) and bound from below by the sBH mass accretion rate. 
While this result is numerically close to that obtained by \cite{Nicholas_runawayBH_2017}, the methods of modeling differ. We use a stronger sBH segregation and allow the growing mass of the IMBH to reshape the distribution within its radius of influence.
\begin{figure*}
    \centering
    \includegraphics[width=1\linewidth]{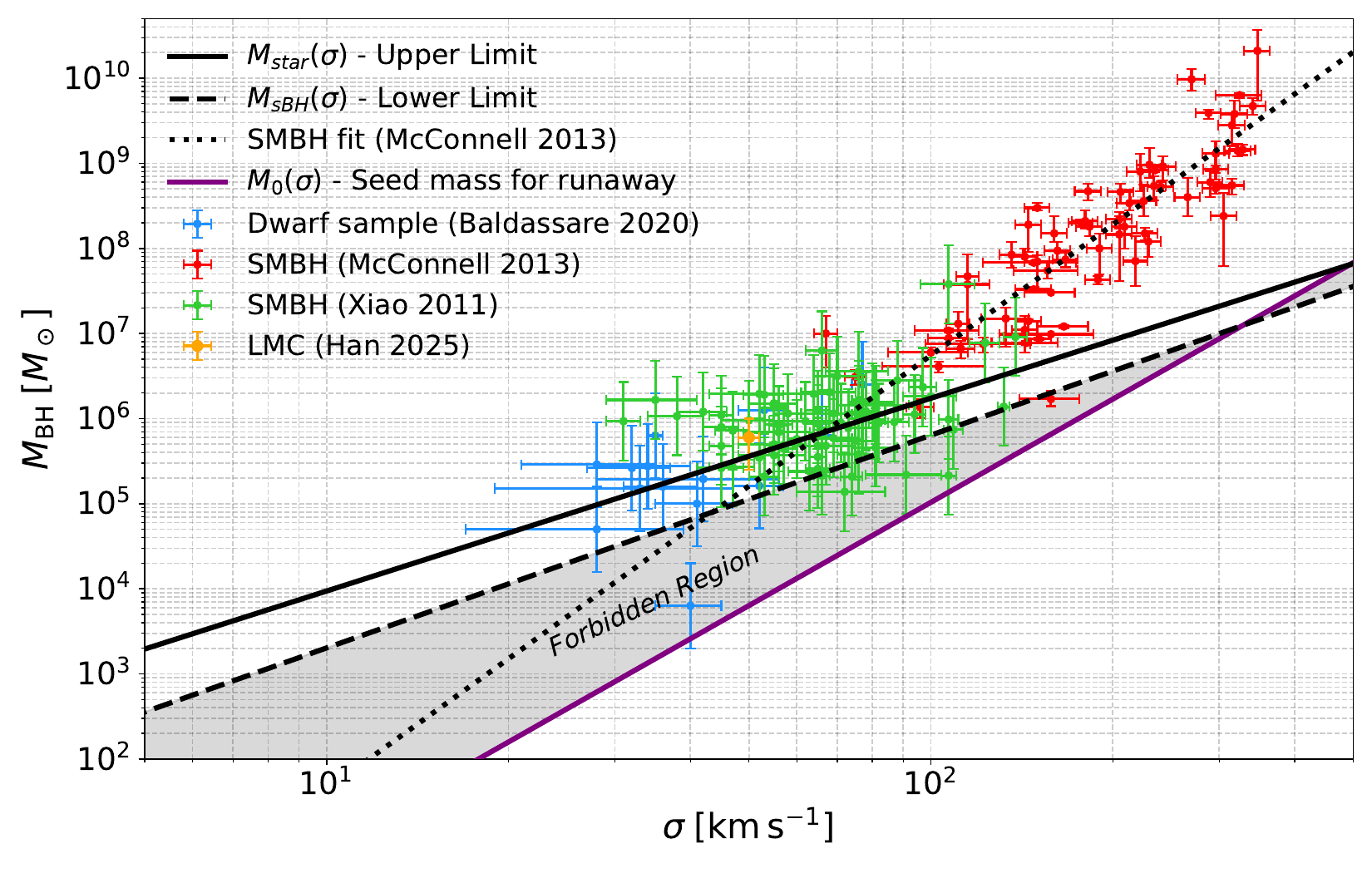}
    \caption{Predicted $M$-$\sigma$ scaling compared to observational samples. The observed relation was taken from \eqref{eq:M_sigma}. We predict a power law change to \(\sigma^{2.26-2.5}\) at \(M\approx10^5M_{\odot}\). Our model does not contradict SMBH samples \citep{Xiao_2011,McConnell_2013,han2025hypervelocitystarstracesupermassive} and fits the dwarf galaxies sample \citep{dwarf_galaxies_M_sigma} better than the extrapolation of the observed relation at high velocity dispersion. The purple line indicates the seed mass required for runaway growth in a core of density $\rho_c = 10^7\mathrm{M_\odot\,{pc}^{-3}}$ and velocity dispersion $\sigma$}
    \label{fig:M-sigma}
\end{figure*}
\section{Growth in a Core Profile}
\label{sec:GC Core}
In this section we consider the growth of small BH seeds for which \(r_h < r_c\). The core mass , which is also the critical MBH mass where \(r_h = r_c\), is
\begin{align}
    M_c &= \frac{\sigma^3}{G\sqrt{\frac{4}{3}\pi G \rho_c}} =\\
    &= 6.79\times10^4M_{\odot}\parfrac{\sigma}{50 \mathrm{km\,s^{-1}}}^{3}\parfrac{\rho_c}{10^7\mathrm{M_\odot\,{pc}^{-3}}}^{14/9} \nonumber.
\end{align}

Since sBH and TDE mass capture rates are comparable, with the TDE mass capture rate being a factor of \(\sim3\) larger  (\ref{eq:sBH_mass},\ref{eq:TDE_mass}), we disregard the sBH mass capture rate in order to get a simple ballpark estimate. Starting with \eqref{eq:TDE_rate_rd} and multiplying by \((\rho_c/\rho_{IS}(r_h))^{14/9}\) to account for the external density change, we get the TDE rate in a constant density core:
\begin{equation}
    \begin{split}
       {\cal R}_{\mathrm{c}}& = K_{\mathrm{c}}\,\,M^{62/27}
               \\&=3.58\times10^{-6}\mathrm{yr^{-1}}\parfrac{M}{10^4M_\odot}^{62/27}\times\\&\parfrac{\sigma}{50 \mathrm{km\,s^{-1}}}^{-47/9}\parfrac{\rho_c}{10^7 \mathrm{M_\odot\,{pc}^{-3}}}^{14/9},
    \end{split}
\end{equation}
where
\begin{equation}
 K_{\mathrm{c}}
        =\frac{2^{53/9}\,\pi^{5/9}}{5^{13/9}\,\ln(\Lambda)^{5/9}}
\,\frac{\rho_c^{14/9}\,r_\star^{5/9} G^{28/9}}{f_\bullet^{1/5}\,m_\bullet^{7/90}\,m_\star^{179/270}\sigma^{47/9}}.\nonumber
\end{equation}
Notice that the growth rate scaling \(M^{62/27}\) corresponds to super-exponential growth, that is, the time to double the mass of the MBH decreases with its increasing mass. Solving the resulting growth equation \(  \dot M= \alpha\,m_{\star}\,K_{\mathrm{c}}\,M^{62/27}\), we get a diverging mass
\begin{equation}
    M(t)=\Bigl[M_0^{-35/27}-\tfrac{35}{27}
                  \alpha m_{\star}  K_{\mathrm{c}}\,t\Bigr]^{-27/35}.
\end{equation} 
This diverging mass is not physical, as it breaks once \(r_h\) reaches the cluster core radius, at which time the MBH mass equals the core's mass $M_c$. Then growth proceeds more slowly according to equations (\ref{eq:sBH_mass},\ref{eq:TDE_mass}).
For a given seed mass \(M_0\) we estimate the growth timescale

\begin{align}
      t\approx\frac{1}{\alpha m_{\star}\,K_{\mathrm{c}}\,M_0^{35/27}},
\end{align}
Using typical NSC core parameters we get
\begin{align}
t\approx5.59Gyr&\parfrac{\rho_c}{10^7\mathrm{M_\odot\,{pc}^{-3}}}^{-14/9}\parfrac{m_{\bullet}}{10\mathrm{M_{\odot}}}^{7/90}\times\\&\parfrac{\sigma}{50 \mathrm{km\,s^{-1}}}^{47/9} \parfrac{M_0}{10^4\mathrm{M_{\odot}}}^{-35/27}\nonumber .
\end{align}
Substituting a galactic lifetime of \(10\mathrm{Gyr}\), we find the minimal seed mass required to ignite this runaway process

\begin{align}
\label{eq:minimal_seed_for_runaway}
M_0 \approx 6400M_{\odot} & \left(\frac{\rho_c}{10^7\mathrm{M_\odot\,{pc}^{-3}}}\right)^{-6/5}\parfrac{m_{\bullet}}{10\mathrm{M_{\odot}}}^{-3/50}\times \nonumber \\
&\left(\frac{\sigma}{50 \mathrm{km\,s^{-1}}}\right)^{141/35}.
\end{align}

This seed mass is substantial, well above the mass of sBHs. Therefore, it is not guaranteed that growth would start. Yet, this seed mass lies below the core mass $M_c$ and well below the extrapolation of the observed $M$-$\sigma$ relation \eqref{eq:M_sigma}.
This implies a "forbidden region" in the $M$-$\sigma$ plane, bounded from below by $M_0$ and from above by our derived capture-driven scaling (\ref{eq:sBH_mass}, \ref{eq:TDE_mass}). Any black hole within this strip $M_0<M<M_\star$ must have accrete mass and therefore has to follow our predicted $M$-$\sigma$ relation  (\ref{eq:sBH_mass}, \ref{eq:TDE_mass}).
Furthermore, the strong dependence of the seed mass on velocity dispersion ($\sigma^{4.03}$) implies that for lower-velocity dispersion systems, the barrier to runaway drops. In the regime of globular clusters, the required seed mass enters the stellar-mass black hole domain. For example, applying the parameters of the dense globular cluster M92 (\(\rho_c = 2\times10^{6}\mathrm{M_\odot\,{pc}^{-3}}\), \(\sigma = 9.1\mathrm{km\,s^{-1}}\); \citealt{Hilker_2019}), we find that a seed of only \(M_0 \approx 46\,M_\odot\) is sufficient. This suggests that in the densest globular clusters, stellar remnants may serve as the progenitors for IMBHs via this channel. Indeed, several known globular clusters such as $\omega$~Cen \citep{prieto2025growingintermediatemassblackhole} and M~15 show hints of central IMBHs.

A seed of sufficient size could be formed in nuclear stellar clusters and dense globular clusters via a collision runaway \citep{2004Natur.428..724P,Nicholas_runawayBH_2017} or hierarchical stellar mass BH mergers \citep{Miller_2002}, then grow rapidly to match \(M_c\) and proceed to grow slowly until it follows  (\ref{eq:sBH_mass}, \ref{eq:TDE_mass}). 
The growth runaway described above could serve as the bridge between the higher mass end of the stellar BH and IMBH regime to the SMBH regime, adding one more piece to the puzzle of MBH forming mechanisms.
\subsection{BH Merger Recoil}
Our model assumes that the growing MBH is retained at the cluster center by dynamical friction. However, during the early growth phase, the anisotropic emission of gravitational waves from sBH mergers imparts a recoil kick to the remnant \citep{Lousto_2011}. If this recoil velocity exceeds the cluster's escape velocity, $v_{\rm esc}$, the seed will be ejected, halting growth.

For non-spinning black holes, the recoil velocity scales with the symmetric mass ratio $\nu$ as $V_{\rm kick} \approx 0.0467\, c \, \nu^2$ \citep{Rom_2022}. Using a conservative escape velocity of  $v_{\rm esc} \approx 10\,\mathrm{km\,s^{-1}}$, retention requires $\nu \lesssim 0.027$. For a characteristic inspiraling mass of $m_\bullet = 10\,M_\odot$, this condition implies that the central black hole is vulnerable to ejection until it exceeds $M \sim 400\,M_\odot$. Notably, this stability threshold is close to the minimum seed mass required for runaway growth (Equation \ref{eq:minimal_seed_for_runaway}); thus, seeds massive enough to trigger the runaway are generally massive enough to survive gravitational recoil.
\section{DISCUSSION \& SUMMARY}
\label{sec:conclusions}
We set out to examine whether sBH capture and TDE accretion can grow massive black holes and thereby flatten the observed $M$-$\sigma$ relation.  
Using a simple two-mass model, we derived closed–form fluxes, and obtained analytic growth laws.
Integration over time yields \(M\propto\sigma^{\beta}\) with $2.26<\beta<2.5$, implying a flattening of the $M$-$\sigma$ slope below \(M\!\sim\!10^5~M_{\odot}\) , the domain where several surveys already hint at increased scatter or a break.  \\
Furthermore, if some clusters contain unusually low core densities, the increased scattering can also be explained as such clusters avoid runaway.
Our mechanism provides a mass floor for central black holes. While other processes (such as gas inflows) may operate and drive the mass higher, capture-driven growth sets the minimum mass that a black hole above the runaway threshold will achieve in a dense cluster environment.
The proposed mechanism is related to that discussed by \citet{Nicholas_runawayBH_2017}, who modeled black hole runaway growth through tidal captures in a time-dependent nuclear star cluster (NSC). Their work follows the dynamical evolution of a NSC, including core heating by binary sBHs and subsequent core collapse. In contrast, our treatment adopts a simplified, static cluster model with a core–isothermal density profile and an embedded, mass-segregated Bahcall–Wolf cusp that evolves in time. This framework allows us to describe stellar and sBH accretion self-consistently and to derive closed-form expressions for the growth rates and present-day masses. Because our model assumes an already collapsed, steady-state core, we need not follow the cluster evolution explicitly.
Compared to their work, we use a modern version of mass segregation and take into account the cusp profile created by the growing MBH. We focus on the expected present day size of the IMBH due to these captures, rather than the runaway stage.
In summary, a purely capture driven channel for black-hole growth explains some initial hints of flattening and increased scatter of the $M$-$\sigma$ relation in the low-\(\sigma\) regime.
\begin{acknowledgments}
The authors would like to thank Barak Rom, Elisha Modelevsky, Eliot Quataert and Tsvi Piran for useful discussions. This research was partially supported by an ISF grant, an NSF/BSF grant, an MOS grant, and a GIF grant.
\end{acknowledgments}
\appendix
\section{One-Mass Model}
\label{sec:oneMass}
The following derivation considers a one-mass model, in this paper we add the needed corrections to model a more realistic two-mass environment. We attach this derivation for completeness.

Around a MBH, objects experience gravitational encounters that cause their angular momenta to randomly diffuse. Over time, some are scattered into an orbit that brings them close enough to the MBH to be captured, this region of phase space is called the \emph{loss cone}. We estimate the rate at which objects are scattered into the loss cone.

An object on a nearly parabolic orbit around a central MBH with a pericenter radius equal to the capture radius \(r_p\approx R_{LC}\) has specific angular momentum
\begin{equation}
  J_{\mathrm{LC}} \approx \sqrt{2GMR_{LC}}\;,
  \label{eq:J_from_R_LC}
\end{equation}
where \(M\) is the central MBH mass.\\
Any object with \(J<J_{LC} \) will be consumed by the MBH once it gets to the pericenter.
Objects with Semi Major Axis (SMA) \(r\) have a characteristic specific angular momentum \(
  J_c(r)=\sqrt{GMr}
\), corresponding to the circular orbit angular momentum. Due to two-body relaxation, angular momentum undergoes diffusion in the two-dimensional plane \((J_x, J_y)\) with an effective diffusion coefficient
\begin{equation}
     D \sim \frac{J_c^2}{\trlx(r)},
\end{equation}
where \(\trlx\) is the local relaxation time \citep{binney2008galactic}.
\subsection{Empty Loss Cone Rate}
Under this approximation objects with \(J<J_{LC}\) capture immediately since the orbital time is short compared to the diffusion timescale. Therefore we can solve the diffusion equation with a loss condition at \(J_{LC}\) and find the diffusive current, which is the loss rate.
Consider the steady-state diffusion equation of the number density in the 2D angular momentum phase space \(f(J)\) where \(J=\sqrt{J_x^2+J_y^2}\)
\begin{equation}
  \frac{1}{J}\frac{d}{dJ}\Bigl(J\,\frac{df}{dJ}\Bigr)=0.
  \label{eq:diffusion}
\end{equation}
The empty loss cone (LC) boundary conditions are:
\begin{enumerate}
  \item The loss condition \(f(J_{\mathrm{LC}})=0\).
  \item At \(J_c\), the distribution reaches its ``background'' value \(f(J_c) = f_0 =  \frac{N(r)}{\pi J_c^2}\),
\end{enumerate}
where the number of objects enclosed within $r$ is \citep{1976ApJ...209..214B} 
\begin{equation}
  N(r)=\frac{16\pi}{5m}\,\rho_h\,r_h^{7/4}\,r^{5/4} 
  \label{eq:Nofr}
\end{equation}
(\(m\) is the average object mass up to radius \(r\), \(\rho_h = \rho_{ext}(r_h)\)). Thus, the distribution is
\begin{equation}
  f(J) = f_0\,\frac{\ln(J/J_{\mathrm{LC}})}{\ln(J_c/J_{\mathrm{LC}})}.
\end{equation}
The supply rate \(\Gamma_{emp}(r)\) into the loss cone under the empty LC approximation is given by the diffusive current
\begin{equation}
  \Gamma _{emp}(r)= -2\pi J\,D\,\left.\frac{df}{dJ}\right|_{J=J_{\mathrm{LC}}}
=  \frac{2N(r)}{\trlx(r)\,\ln(J_c/J_{\mathrm{LC}})}.
  \label{eq:flux}
\end{equation}
For the BW profile relaxation time is \citep{Spitzer1987, binney2008galactic} 
\begin{equation}
  \trlx(r) = \frac{P(r)}{\ln \Lambda}\left(\frac{M}{m}\right)^2\frac{1}{N(r)},
  \label{eq:Trel}
\end{equation}
where \(P(r)\) is the orbital period and \(\ln\Lambda \approx10\) is the Coulomb logarithm \citep{Spitzer1987, binney2008galactic} . We get \( \Gamma_{emp}(r) \propto r\), indicating that most stars are consumed from the outermost empty LC region.
\subsection{Full Loss Cone Rate}
In the full loss cone regime the angular momentum change over one orbit \(\Delta J\) is large compared to \(J_{LC}\). Therefore the loss cone has minimal effect on the distribution and objects are distributed uniformly in the \((J_x,J_y)\) space \citep{Shapiro1977ApJ,1976MNRAS.176..633F}.  The critical radius that separates both regimes is \(r_d\) which satisfies \(\Delta J(r_d)= J_{LC}\)
\begin{equation}
\Delta J(r) \sim J_c(r)\sqrt{\frac{P(r)}{\tau_{\rm rlx}(r)}},
\end{equation}
Solving for \(r_d\) we get
\begin{equation}
  r_d \;=\;\left(\frac{5M}{16\pi \ln{\Lambda}} \frac{M}{m} \frac{R_{LC} }{\rho_h r_h^{7/4}}\right)^{4/9}.
  \label{eq:rd_GC}
\end{equation}

The number of objects with \(J<J_{LC}\) is \(N(r)\cdot\parfrac{J_{\rm LC}}{J_c(r)}^2\) , so the supply rate in the full loss cone regime of objects originating from radius \(r\) is
\begin{equation}
\Gamma_{\rm full}(r) = N(r)\cdot\parfrac{J_{\rm LC}}{J_c(r)}^2 \frac{1}{P(r)}.
\end{equation}
Therefore  \(\Gamma_{full}(r) \propto r^{-5/4}\)  so most full loss cone events originate from \(r_d\), the lowest radius that satisfies full LC conditions. Under empty LC conditions, most events originate from the highest radius which is also \(r_d\) for \(r_d<r_h\) . Since \(\Gamma_{full}(r_d) = \Gamma_{emp}(r_d)\) we estimate the overall rate to be
\begin{equation}
\label{eq:full_LC_flux}
{\cal R} = \Gamma_{emp}(r_d).
 \end{equation}
If \(r_d>r_h\) there are no objects in the full LC regime and the rate is
\begin{equation}
 {\cal R} = \Gamma_{emp}(r_h).
\end{equation}
Since \( \Gamma_{emp}(r)\) is an increasing function
\begin{equation}
{\cal R} = \min(\Gamma_{emp}(r_h),\;\Gamma_{emp}(r_d)).
\end{equation}
Substituting we get
\begin{align}
\label{eq:transition_rate}
    \Gamma_{emp}(r_d)  &=\frac{2^{56/9}\,\pi^{5/9}}{5^{14/9}\ln^{4/9}\Lambda}\times\\&
         \times \frac{\rho_h^{14/9}\,G^{29/9}\,R_{LC}^{4/9}\,M^{19/9}}
               {\sigma^{49/9}\,m^{4/9}},\nonumber
\end{align}
and
\begin{equation}
    \label{eq:rhrate}
    \Gamma_{emp}(r_h)=\frac{256\pi\rho_h^2G^5M^3}{25\sigma^9}.
\end{equation}
To get the rate, we substitute the external density profile and the loss cone radius. Note that using \eqref{eq:J_from_R_LC} we can replace $R_{LC}$ with $J_{LC}$.
\section{Derivation of Eddington Limited TDE Efficiency }
\label{app:EddEffderivation}
In this derivation we estimate \(\alpha_{Edd}\), the efficiency of Eddington limited TDEs. We use it as a lower bound on \(\alpha\). We use the duration of the super-Eddington phase as given by \cite{Quataert_Strubbe_TDE2009}. Inflow onto the MBH is Eddington limited at \(\dot M_{\rm Edd}=L_{\rm Edd}/(\eta c^2)\propto M_\bullet/\eta\) and then becomes sub–Eddington and falls following \(t^{-5/3}\). The fraction ultimately accreted can be found by integrating \(\dot M(t)\) and dividing by the bound mass
\begin{align}
M_{acc}&=  \dot M_{\rm Edd} \,(t_{\rm Edd}-t_{\rm fb})
+ \int_{t_{\rm Edd}}^\infty \dot M_{\rm fb}(t)\,dt
\\&\approx\dot M_{\rm Edd}\,t_{\rm Edd}+ \frac{m_\star}{2}\left(\frac{t_{\rm Edd}}{t_{\rm fb}}\right)^{-2/3}.
\end{align}
Substituting values for \(t_{\rm Edd},\dot M_{\rm Edd},t_{\rm fb}\) from \cite{Quataert_Strubbe_TDE2009} and \(m_\star= M_\odot , r_\star=r_\odot\), \(\eta=0.1\) we get
\[
\alpha_{Edd} = 0.021\,
\parfrac{M}{10^5M_\odot}^{3/5}\,\parfrac{m_\star}{M_\odot}^{-4/5}\,\parfrac{r_\star}{R_\odot}^{3/5}.
\]
This assumes that super–Eddington supply is not stored but largely expelled.\\
\bibliography{sample7}
\bibliographystyle{aasjournal}
\end{document}